\documentclass[12pt]{article}
\usepackage{amssymb,latexsym}
\usepackage[mathscr]{eucal}
\usepackage{amsmath,amssymb}
\usepackage{latexsym}
\usepackage{bbm}
\usepackage{bm}

\newcommand{\beq}{\begin{equation}}
\newcommand{\eeq}{\end{equation}}
\newcommand{\beqn}{\begin{eqnarray}}
\newcommand{\eeqn}{\end{eqnarray}}

\newcommand{\nn}{\nonumber}

\numberwithin{equation}{section}

\newcommand{\Z}{\mathbbm{Z}}

\newcommand{\be}{\begin{equation}}
\newcommand{\ee}{\end{equation}}
\newcommand{\ba}{\begin{eqnarray}}
\newcommand{\ea}{\end{eqnarray}}
\newcommand{\bdm}{\begin{displaymath}}
\newcommand{\edm}{\end{displaymath}}

\def\b{\beta}
\def\a{\alpha}
\def\g{\gamma}

\newcommand{\ie}{{\it i.e.\ }}
\newcommand{\eg}{{\it e.g.\ }}

\DeclareMathAlphabet{\mathpzc}{OT1}{pzc}{m}{it}
\def\bea{\begin{eqnarray}}
\def\eea{\end{eqnarray}}
\def\beas{\begin{eqnarray*}}
\def\eeas{\end{eqnarray*}}
\def\sla{\raise.15ex\hbox{$/$}\kern-.57em}

\def\bea{\begin{eqnarray}}
\def\eea{\end{eqnarray}}

\def\sla{\raise.15ex\hbox{$/$}\kern-.57em}
\def\ie{{\it i.e.}~}
\def\eg{{\it e.g.}~}
\def\ap{{\alpha^\prime}}

\def\a{\alpha}
\def\b{\beta}
\def\g{\gamma}

\def\cA{{\cal A}}

\def\cI{{\cal I}}

\def\cM{{\cal M}}
\def\cN{{\cal N}}

\def\cQ{{\cal Q}}

\def\cT{{\cal T}}

\def\cX{{\cal X}}

\begin{document}
\begin{titlepage}
\begin{flushright}
CERN-PH-TH/2008/095 \\
{ROM2F/2008/14}\\
\end{flushright}
\vskip 0.5cm

\begin{center}
{\large \sc  Bound-states of D-branes in \\
L-R asymmetric superstring vacua}\\ \vspace{1.0cm}
{\bf Massimo Bianchi$^{1,2}$}\\
\vskip 0.5cm $^1${\sl Physics Department, Theory Unit, CERN \\ CH
1211, Geneva 23,
Switzerland }\\
$^2${\sl Dipartimento di Fisica \& Sezione INFN \\
 Universit\`a di Roma ``Tor Vergata''\\
Via della Ricerca Scientifica, 00133 Roma, Italy}\\

\end{center}
\vskip 2.0cm
\begin{center}
{\large \bf Abstract}
\end{center}

We discuss bound-states of D-branes in truly L-R asymmetric and
thus non-geometric Type II vacuum configurations with extended
supersymmetry. We argue for their stability as a result of
residual supersymmetry and coupling to R-R potentials surviving in
the massless spectrum. We then identify the open string
excitations of these L-R asymmetric BPS D-branes. Finally, we
briefly comment on possible applications and extensions.

\vfill
\end{titlepage}

\section{Introduction}

D-branes represent the best known class of non-perturbative states
in String Theory since they admit a world-sheet description in
terms of open strings \cite{Dai:1989ua, Polchinski:1995mt}. They
couple minimally to R-R potentials \cite{Polchinski:1995mt,
Bianchi:1991eu} and break half of the original supersymmetries of
Type II superstrings.

D-branes have been more or less explicitly considered in a variety
of geometric \cite{Angelantonj:2002ct, Uranga:2003pz,
Kiritsis:2003mc, Blumenhagen:2005mu, Blumenhagen:2006ci} and
non-geometric \cite{Bianchi:1991rd, Pradisi:1995qy,
Bianchi:1997gt, Brunner:1999fj, Bianchi:1999uq,
Blumenhagen:2000fp, Angelantonj:2000xf, Gutperle:2000bf}
 contexts whereby the parent Type II
theory admits some involution exchanging Left and Right movers
\cite{Sagnotti:1987tw, Bianchi:1988fr, Pradisi:1988xd,
Bianchi:1989du, Horava:1989vt, Bianchi:1990tb, Bianchi:1990yu,
Gimon:1996rq}.

Very recently the existence and properties of D-branes in
genuinely L-R asymmetric and thus non geometric Type II vacuum
configurations has started being investigated
\cite{Lawrence:2006ma, Kawai:2007qd}. These include asymmetric
orbifolds \cite{Narain:1986qm}, free fermion constructions
\cite{Kawai:1986ah, Antoniadis:1986rn, Antoniadis:1987wp} and
covariant lattices \cite{Lerche:1986cx}. As pioneered by Ferrara
and Kounnas \cite{Ferrara:1989nm}, these constructions allow the
embedding of $D=4$ extended supergravities with $\cN = \cN_{_L} +
\cN_{_{_R}}$ supersymmetries in Type II superstrings. While all
$\cN$ even cases except $\cN =6$ admit, but by no means require,
L-R symmetric descriptions, all odd cases and $\cN=6$ require non
L-R symmetric descriptions\footnote{Excluding for the time being
the possibility of enhanced supersymmetry (extra massless
gravitini) from twisted sectors}.  Up to $\cN_{_L} \leftrightarrow
\cN_{_R}$ interchange, the list of possibilities
reads\footnote{$\cN=7$ is equivalent to $\cN=8$.} \bea \cN = 8
\quad &\leftrightarrow&\quad \cN_{_L} = 4 \ , \ \cN_{_R} = 4 \\
\cN = 6 \quad &\leftrightarrow& \quad \cN_{_L} = 2 \ , \ \cN_{_R}
= 4 \\ \cN = 5 \quad &\leftrightarrow& \quad \cN_{_L} = 1 \ , \
\cN_{_R} = 4 \\ \cN = 4 \quad &\leftrightarrow& \quad \cN_{_L} = 2
\ , \ \cN_{_R} = 2 \quad {\bf or}\quad \cN_{_L} = 0 \ , \ \cN_{_R}
= 4 \\ \cN = 3 \quad &\leftrightarrow& \quad \cN_{_L} = 1 \ , \
\cN_{_R} = 2 \\ \cN = 2 \quad &\leftrightarrow& \quad \cN_{_L} = 1
\ , \ \cN_{_R} = 1 \quad {\bf or}\quad \cN_{_L} = 0 \ , \ \cN_{_R}
= 2 \\ \cN = 1 \quad &\leftrightarrow& \quad \cN_{_L} = 0 \ , \
\cN_{_R} = 1 \eea

When $\cN_{_L}=0$ or $\cN_{_R}=0$, all R-R fields are massive and
we expect no BPS bound state of D-branes. Stable non-BPS D-branes
with torsional K-theory charges may exist though
\cite{Witten:1998cd, Sen:1999mg, Lerda:1999um}. In all other
cases, massless R-R fields survive that \emph{must} couple to some
kind of D-branes. Since some extended supergravities can be
obtained by spontaneous supersymmetry breaking in the presence of
internal closed string fluxes \cite{Frey:2002hf,
Dall'Agata:2005ff}, given the uniqueness of the theories with $\cN
\ge 5$ and the rather rigid structure of the theories with $\cN =
3,4$, it is tempting to conjecture some form of duality between
(D-branes in) non-geometric backgrounds and (D-branes in)
geometric flux compactifications. In fact one can turn the
argument the other way around. Given our limited knowledge as how
to quantize string theory in flux backgrounds
\cite{Berkovits:1999im, Berkovits:2001tg, Linch:2006ig,
Linch:2008rw}, one can use the non geometric world-sheet
construction as an equivalent `dual' definition of the latter.
Indeed duality between geometric fluxes ($H_{ijk}$ 3-form and
$T_{ij}{}^k$ torsion) and non geometric ones ($Q_i{}^{jk}$ and
$R^{ijk}$) has been proposed and supported by some evidence
\cite{Dabholkar:2002sy, Dabholkar:2005ve, Hull:2006tp,
Hull:2006va, Shelton:2006fd, Grange:2006es, Berman:2007vi,
Villadoro:2007yq, Berman:2007xn, Villadoro:2007tb, Wecht:2007wu,
Hull:2007jy, Wecht:2007wu, Dall'Agata:2007sr, Berman:2007yf}.

Clearly, once the correspondence has been established in highly
supersymmetric contexts as those we focus on in this note, it is
natural to argue that it hold in lower or non supersymmetric
configurations, albeit with massless R-R fields. The fate of L-R
asymmetric D-branes in the absence of massless R-R fields is an
interesting problem that we will not address here.

The plan of this note is as follows. In section 2 we discuss the
simplest non-trivial case, $\cN = 6$ supergravity, and identify
the surviving massless R-R vectors and the L-R asymmetric D-branes
they couple to. We extend the analysis to other L-R asymmetric
realizations of extended supergravities in Section 3. Then, in
Section 4, we construct the invariant boundary states and discuss
the relevant open string excitations. Section 5 contains our
conclusions and comments.

\section{The $\cN = 6$ case with $\cN_{_L}=2$ and $\cN_{_R}=4$}
\label{N6}

In order to illustrate our point let us start with the simplest
non trivial case, $\cN =6$ supergravity with 24 supercharges
\cite{Ferrara:1989nm, Villadoro:2004ec}. The highest dimension
where the classical theory can be defined is $D=6$. However the
resulting $\cN =(2,1)$ supergravity is anomalous and thus
inconsistent at the quantum level
\cite{D'Auria:1997cz}\footnote{String theory prevents quantum
inconsistencies thanks to the presence of new massless `chiral'
(twisted) states whenever modular invariance or tadpole
cancellation is imposed.}. So we are led to consider $D=5$. One
starts with a toroidal compactification and quotients it by a
chiral $\Z_2$ twist of the L-movers (`T-duality' on four internal
directions) \be X^i_{_L} \rightarrow - X^i_{_L} \quad , \quad
\Psi^i_{_L} \rightarrow -\Psi^i_{_L} \quad , \quad  i=6,7,8,9
\label{Z2twist}\ee accompanied by an order two shift compatible
with modular invariance in order to make twisted states massive.
In Type II theories, chiral supersymmetric twists such as
(\ref{Z2twist}) are anyway level-matched by themselves. For
definiteness, we consider the Type IIB case. In the untwisted
sector the one-loop torus partition function reads \be \cT_u =
{1\over 2} \left\{(Q_o + Q_v) \bar{Q} \Lambda_{5,5}[^0_0] + (Q_o -
Q_v)(X_o - X_v) \bar{Q} \Lambda_{1,5}[^0_1] \right\} \ee where
$X_o - X_v = 4 \eta^2/\theta_2^2$ encodes the effect of  the
$\Z_2$ projection on four internal L-moving bosons, \bea
\Lambda_{l,r}[^a_b] = \sum_{p_{_L}, p_{_R}} e^{i\pi [a_{_L} p_{_L}
- a_{_R} p_{_R}]} q^{{1\over 2} (p_{_L} + {1\over 2} b_{_L})^2}
\bar{q}^{{1\over 2} (p_{_R} + {1\over 2} b_{_R})^2} \eea are
(shifted) Lorentzian lattice sums of signature $(l,r)$ and $Q=V_8
- S_8$, $Q_o = V_4 O_4 - S_4 S_4$, $Q_v = O_4 V_4 - C_4 C_4$ with
$O_n, V_n, S_n, C_n$ representing the characters of $SO(n)$ at
level $\kappa=1$\footnote{For $n$ odd $S_n$ coincides with $C_n$
and will be denoted by $\Sigma_n$.}. By modular transformations
$S$ and then $T$ one finds the twisted sector \be \cT_t = {1\over
2} \left\{(Q_s + Q_c)(X_s + X_c) \bar{Q} \Lambda_{1,5}[^1_0] +
(Q_s - Q_c)(X_s - X_c) \bar{Q} \Lambda_{1,5}[^1_1] \right\}\ee
where $X_s + X_c = 4\eta^2/\theta_4^2$, $X_s - X_c =
4\eta^2/\theta_3^2$, $Q_s = O_4 S_4 - C_4 O_4$ (`massless'), $Q_c
= V_4 C_4 - S_4 V_4$ (`massive') a.

Due to the shift, the massless spectrum receives contribution only
from the untwisted sector. In $D=5$ notation with $SO(3)$ little
group one finds  \bea &&(V_3 + O_3 - 2\Sigma_3) \times (\bar{V}_3
+ 5 \bar{O}_3 - 4 \bar{\Sigma}_3)
\rightarrow \\
&&(g + b_2 + \phi)_{_{NS-NS}} + 6 A_{_{NS-NS}} + 5 \phi_{_{NS-NS}}
+ 8 A_{_{R-R}} + 8 \phi_{_{R-R}} - {Fermi} \nn  \eea

The hidden non-compact symmetry is $SU^*(6)$. After dualizing all
masseless 2-forms into vectors,  the ${\bf 15} =7_{_{NS-NS}}+
8_{_{_{R-R}}}$ vectors transform according to the antisymmetric
tensor of $SU^*(6)$. The $14=1_{_{NS-NS}}+5_{_{NS-NS}}+8_{_{R-R}}$
scalar moduli parameterize the space $\cM^{D=5}_{\cN = 6} =
SU^*(6)/Sp(6) $.

Reducing to $D=4$ on another circle with or without further
shifts, the massless spectrum is given by \bea &&(V_2 + 2 O_2 -
2S_2 - 2 C_2) \times (\bar{V}_2 + 6 \bar{O}_2 - 4 \bar{S}_2 - 4
\bar{C}_2)
\rightarrow \\
&&(g + b + \phi)_{_{NS-NS}} + 8 A_{_{NS-NS}} + 12 \phi_{_{NS-NS}}
+ 8 A_{_{R-R}} + 16 \phi_{_{R-R}} - {Fermi} \nn  \eea

The hidden non-compact symmetry is $SO^*(12)$.  After dualizing
all masseless 2-forms into axions, the
$30=2_{_{NS-NS}}+12_{_{NS-NS}}+16_{_{R-R}}$ scalar moduli
parameterize the space $ \cM^{D=4}_{\cN = 6} = SO^*(12)/U(6) $.
The $16=8_{_{NS-NS}}+8_{_{R-R}}$ vectors together with their
magnetic duals transform according to the ${\bf 32}$ dimensional
chiral spinor representation of $SO^*(12)$.

In order to identify the conserved charges coupling to the
surviving R-R and NS-NS graviphotons, it is convenient to first
consider maximal $\cN = 8$ supergravity in $D=4$ obtained by
compactification on $T^6$. The 12 NS-NS graviphotons couple to
windings and KK momenta. Their magnetic duals to wrapped
NS5-branes (H-monoples) and KK monopoles. The 32 R-R graviphotons
(including magnetic duals) couple to (6) D1-, (6) D5- and (20)
D3-branes.

The chiral $\Z_2$ projection from $\cN = 8$ to $\cN=6$ eliminates
4 NS-NS vectors coupling to $p_{_L}$ along the 4 twisted
directions (so that $p^i_{_L} = 0$ implies $n^i=m^i$ along the
directions $i=6,7,8,9$) and 8 R-R vectors thus `identifying'
different kinds of D-branes. Our aim is to make this statement
more precise, \ie to identify the bound states of D-branes allowed
by the `T-duality' quotient. Our proposal for the surviving
$16_{_{R-R}}=2_{(1|5)} + 4_{(1|3)} + 6_{(3|3)} + 4_{(5|3)}$
D-brane charges is\footnote{We henceforth use an intuitive
notation whereby the subscript indicates which D-branes appear in
the bound-state carrying a particular R-R charge.}
 \be q_1^{a} + {1\over 4!}\varepsilon_{ijkl} q_5^{aijkl}
\quad , \quad q_1^i + {1\over 3!}\varepsilon^i{}_{jkl} q_3^{jkl}
\quad , \quad q_3^{aij} + {1\over 2!} \varepsilon^{ij}{}_{kl}
q_3^{akl} \quad , \quad q_5^{abijk} + \varepsilon^{ijk}{}_l
q_3^{abl} \label{N6RRcharges}\ee where $q_p^{I_1...I_p}$ denote
the `elementary' Dp-brane charges.

In order to give further support to the above identification of
R-R charges, we would like to show that a bound state of a D5
wrapped around the 4 twisted directions and one of the two
untwisted directions and a D1 wrapped around the same circle
preserve 1/3 of the susy of the background. The unbroken susy of
$\cN = 6$ are the ones satisfying \be \cQ_{_L} = \Gamma_{6789}
\cQ_{_L} \label{susybkg}\ee with no conditions on $\cQ_{_R}$. For
D5 wrapped along the twisted $T^4$ and one of the two circles, \eg
the one along the $4^{th}$ direction, the condition is \be
\cQ_{_R} = \Gamma_{04}\Gamma_{6789} \cQ_{_L}
 \ee using (\ref{susybkg}) one finds
 \be \cQ_{_R} = \Gamma_{04} \cQ_{_L}
 \ee
 which is nothing but the condition for the supersymmetry preserved by a D1
 along the `untwisted' direction of the D5. Adding the D1 does not
 reduce supersymmetry any further: the bound-state is 1/3 BPS wrt the
unbroken supersymmetry in the L-R asymmetric Type II background.
It preserves 8 supercharges out of 24 surviving supercharges,
since the 8 $\cQ_{_L}$ completely determine the $\cQ_{_R}$.
Similar considerations apply to the other surviving bound states
of D-branes. Each one preserves 1/3 of the 24 SUSY charges of $\cN
= 6$ supergravity.

 A slightly different analysis applies to the BPS states carrying
charges in the NS-NS sector. The
$16_{_{NS-NS}}=8^e_{(1|1)}+8^m_{(5|5)}$  surviving charges are\be
m_1^{a} \quad , \quad n_1^{a} \quad , \quad p_{_R}^{i} = m_1^{i} +
n_1^{i} \quad ; \quad m_{5a} \quad , \quad n_{5a} \quad ,
\quad\hat{P}_{Ri} = m_{5i} + n_{5i} \label{N6NScharges}\ee where
$n^I_1$ and $m^I_1$ denote windings and KK momenta with
$I=(a,i)=4,...,9$ and $a=4,5$, $i=6,..9$, while $n_{I,5}$ and
$m_{I,5}$ denote H-monopoles (wrapped 5-branes) and KK momenta. In
particular the
 two gravitini that together with their superpartners are rendered massive by the freely
 acting $\Z_2$ projection form a complex 1/2 BPS
 multiplet with mass equal to the KK momentum for $R_5 > \ap$.

 There are many other superstring realizations of $\cN = 6$ supergravity in
 $D=4$. Given the uniqueness of the low-energy theory, they all
 share the same massless spectrum. One possibility is to break
 half of the L-moving supersymmetries by means of a $\Z_n$
 chiral projection acting on 4 supercoordinates as
\be (Z^1, Z^2)_{_L} \rightarrow (\omega Z^1, \omega^{-1} Z^2)_{_L}
\quad , \quad (\Psi^1, \Psi^2)_{_L} \rightarrow (\omega \Psi^1,
\omega^{-1} \Psi^2)_{_L} \ee with $\omega^n =1$. In order to avoid
massless twist with an order $n$ shift along the `untwisted'
directions $(Z^3_{_L};Z_{_R}^i)$. The surviving NS-NS charges (8
electric and 8 magnetic) are as before, see (\ref{N6NScharges}).
The surviving R-R charges (16 including both electric and
magnetic, since a net separation cannot be made for them) are less
intuitive to visualize. Later on we will offer a boundary state
description. For the moment suffice it to say that they are bound
states of different kinds of D-branes obtained one from the other
by the action of the $\Z_n$ twists (chiral rotations) and shifts
(non-geometric translations).

Yet another realization of $\cN = 6$ supergravity in $D=4$ has
been proposed in \cite{Dabholkar:1998kv}. It corresponds to a Type
II compactification on the maximal torus of $SU(3)^3$ with chiral
$\Z_3$ projection and no shift. The untwisted sector yields
$\cN=5$ supergravity while the twisted sector produces the extra
massless gravitino multiplet to complete the spectrum of $\cN=6$
supergravity. In this case, the untwisted sector produces only 6
NS-NS and 4 R-R graviphotons together with 8 NS-NS and as many R-R
(pseudo)scalars. The missing massless states are contributed by
the twisted sector and its conjugate. The 2 twisted NS-NS and 4
twisted R-R graviphotons couple to somewhat exotic charges. The
former to twisted L-moving strings and 5-branes and the latter to
bound states of L-R asymmetric `fractional' D-branes.

 \section{Bound-states of D-branes and R-R Charges}

We would now like to extend the previous analysis for $\cN =6$ to
lower supersymmetric cases in $D=4$. For simplicity, we start with
models obtained by successive $\Z_2$ chiral projections. Later on
we will describe how to deal with $\Z_n$ chiral twists and order
$n$ shifts with $n\neq 2$. As already mentioned, we will not
consider here cases with $\cN_{_L}=0$ or $\cN_{_R} = 0$, typically
but not necessarily involving $(-)^{F_{L/R}}$ projections, since
no massless R-R fields survive in these cases.

\subsection{$\cN =5=1_{_L} + 4_{_R}$ case}

For $\cN =5=1_{_L}+4_{_R}$ using $Z^L_2\times Z^L_2$ which acts by
T-duality along $T^4_{6789}$ and $T^4_{4589}$ combined with order
two shifts, that eliminate massless twisted states, it is easy to
see that the surviving $12_{_{NS-NS}} =6^e_{(1|1)}+6^m_{(5|5)}$
charges in the NS-NS sector are \be p_{_R}^{I} = m_1^{I} + n_1^{I}
\quad ; \quad \hat{P}_{RI} = m_{5I} + n_{5I} \ee In the R-R sector
one finds $8_{_{R-R}} =6_{(1533)} + 2_{(3333)}$ charges \be
q^I_{(1335)}= q_1^{I} + {1\over 4!}\varepsilon_{i_Ij_Ik_Il_I}
q_5^{Ii_Ij_Ik_Il_I} + {1\over 3!}\varepsilon^I{}_{J,K'L'}
q_3^{JK'L'} + {1\over 3!}\varepsilon^I{}_{J,K"L"} q_3^{JK"L"} \ee
where $i_I,j_I,k_I,l_I$ run over the four directions orthogonal to
$T^2_I$ while $K',L'$ and $K", L"$ run over the two sets of two
directions orthogonal to $T^2_I$ and \be q_{(3333)}^{I_1I_2I_3} =
q_{3}^{I_1I_2I_3} + {1\over 2!}\varepsilon^{I_2I_3}{}_{J_2J_3}
q_3^{I_1J_2J_3} + {1\over 2!}\varepsilon^{I_3I_1}{}_{J_3J_1}
q_3^{J1I_2J_3} + {1\over 2!}\varepsilon^{I_1I_2}{}_{J_1J_2}
q_3^{J_1J_2I_3}\ee

Bound states of D-branes carrying the above charges are 1/5 BPS
since they preserve 4 supercharges out of the 20 supercharges
present in the background, \ie those satisfying with $\cQ_{_L} =
\Gamma_{6789}\cQ_{_L}$ and $\cQ_{_L} = \Gamma_{4567} \cQ_{_L}$.
For instance for the bound state with charge $q^4_{(1335)}$ the 4
residual supercharges are those satisfying $\cQ_{_R} = \Gamma_{04}
\cQ_{_L}$.

As in the $\cN = 6$ case, a different analysis applies to BPS
states carrying KK momenta or windings or their magnetic duals.
However, at variant with the $\cN = 6$, the three massive
gravitini cannot form a single complex 2/5 BPS multiplet. One of
them, together with its superpartners, should combined with string
states which are degenerate in mass at the special rational point
in the moduli space where the chiral $\Z_2\times \Z_2$ projection
is allowed.

In \cite{Ferrara:1989nm}, ``minimal'' $\cN = 5$ superstring
solutions have been classified into four classes which correspond
to different choices of the basis sets of free fermions or
inequivalent choices of shifts in the orbifold language.

Due to the uniqueness of $\cN = 5$ supergravity in $D=4$, all
models have the same massless spectrum, contributed by the
untwisted sector. In addition to the graviton $g_{\mu\nu}$ and 5
gravitini $\psi_\mu$, one has 10 graviphotons $A_\mu$, 11 dilatini
$\chi$ and 10 scalars $\phi$. The latter parameterize
$\cM^{D=4}_{\cN = 5} = SU(5,1)/U(5)$. The graviphotons together
with their magnetic duals transform according to the ${\bf 20}$
complex (3-index totally antisymmetric tensor) representation of
$SU(5,1)$.

It would be interesting to explore how different massive spectra
could affect higher derivative terms in the superstring effective
action. Notice that the 8 R-R axions and the NS-NS axion, dual to
$b_{\mu\nu}$ decouple from perturbative amplitudes. Only
non-perturbative effects, due to `non-geometric brane wrappings',
could induce dependence on these fields as well as on the dilaton.

Two alternative superstring constructions with $\cN = 5$
supergravity in the massless spectrum have been proposed in
\cite{Dabholkar:1998kv}. The first consists in a $\Z_7$ asymmetric
orbifold of the maximal tours of $SU(7)$. A twist (chiral
rotation) of the L-movers $\theta_{_L} = (\omega_7,\omega_7^2,
\omega_7^4)$ is accompanied by a shift (chiral translation) of the
R-movers such that $7\sigma_{_R} =(1,2,-3,0,0,0,0)$\footnote{Using
a 7-dim notation with $\sum_i x_i = 0$.}. The second consists in a
$\Z_3$ asymmetric orbifold of the maximal tours of $SU(3)^3$. A
twist (chiral rotation) of the L-movers $\theta_{_L} =
(\omega_3,\omega_3, \omega_3)$ is accompanied by a shift (chiral
translation) of the R-movers such as $3\sigma_{_R} =(1,-1,0;
1,-1,0;1,-1,0)$\footnote{Using a 9-dim notation with $\sum_i x^I_i
= 0$ for $I=1,2,3$.}.

In general one expects to get $\cN = 5$ supergravity by means of a
chiral $\Z_n$ twist $\theta_{_L} = (\omega_n^a,\omega_n^b,
\omega_n^c)$ with $a+b+c = 0$ (mod $n$) on a suitable (Lie
algebra) lattice combined with a R shift, satisfying level
matching and not belonging to $I^*$, the dual to the lattice
invariant under $\theta_{_L}$, to avoid massless twisted sectors
that can contribute extra gravitini unless specific choices of
phases (discrete torsions) are made \cite{Dolivet:2007sz}.

\subsection{$\cN =4=2_{_L} + 2_{_R}$ case}

The $\cN =4=2_{_L} + 2_{_R}$ case allows both geometric (L-R
symmetric) and non-geometric (truly L-R asymmetric) descriptions.
In particular one can start by considering the combined effect of
two T-duality projections $\Z_2^L \times \Z_2^R$ on $T^4$.
Depending on the choice of discrete torsion $\epsilon = \pm$ and
shifts $\sigma_V$ one can have various cases. In the absence of
shifts, the model, though non-geometric, turns out to be L-R
symmetric \cite{Bianchi:1999uq}. For $\epsilon = +$, the Type IIB
compactification enjoys $\cN = (2, 2)$ `enhanced' supersymmetry
since massless gravitino multiplets appear in the twisted sectors.
The resulting non-compact symmetry is $SO(5,5)$. For $\epsilon =
-$ the Type IIB compactification enjoys $\cN = (2, 0)$
supersymmetry. In addition to the supergravity multiplet (with 5
self-dual antisymmetric tensors), one has  21 $\cN = (2, 0)$
tensor multiplets, 5 from the untwisted sector and 16 from the
twisted sectors. The hidden non-compact symmetry of the model is
$SO(5,21)$.  Though non-geometric this compactification is
topologically equivalent to a compactification on K3. The torus
partition function involves the `diagonal' modular invariant
rather than the `charge-conjugation' modular invariant
\cite{Bianchi:1999uq, Blumenhagen:2000fp}.

Reducing the model with $\epsilon = +$ to $D=4$, one finds 4
untwisted NS-NS charges $n_1^a$, $m_1^a$ and their magnetic duals
$n_5^a$, $m_5^a$. In the untwisted sector one finds only
$8_{_{R-R}} =2_{(1|5)} + 6_{(3|3)}$ T-duality invariant R-R
charges  charges \be q_1^{a} + {1\over 4!}\varepsilon_{ijkl}
q_5^{aijkl} \quad , \quad q_3^{aij} + {1\over 2!}
\varepsilon^{ij}{}_{kl} q_3^{akl} \ee

In the absence of shifts, there are $16_{_{_{R-R}}}$ additional
R-R charges from the twisted sectors $q_3^f$. Including shifts,
one can eliminate massless states in the twisted sectors and
render the background genuinely asymmetric under the exchange of
Left and Right movers.

Many more L-R asymmetric $\cN =4=2_{_L} + 2_{_R}$ models can be
constructed \cite{Ferrara:1989nm}. A systematic analysis is beyond
the scope of this note.

\subsection{$\cN =3=1_{_L} + 2_{_R}$ case}

The simplest $\cN=3$ model has 3 matter vector multiplets and can
be constructed in two steps \cite{Ferrara:1989nm}.

The first steps consists in a `geometric' $\Z_2$ freely acting
orbifold (locally equivalent to $K3\times T^2$). The $\Z_2$ action
combines a twist breaking $\cN=8=4_{_L} + 4_{_R}$ to $\cN=4=2_{_L}
+ 2_{_R}$ and a shift preventing new massless states from
appearing in the twisted sector. The resulting massless spectrum
consists in $\cN = 4 $ supergravity coupled to 6 matter vector
multiplets. The scalar manifold is \be \cM_{\cN = 4} = {SO(6, 6)
\over SO(6)\times SO(6)} \times {SL(2)\over U(1)}\ee Using the
notation introduced in Sect.~2, the one-loop partition partition
in the untwisted sector reads \be \cT^{\cN=4}_u = {1\over 2} \{
|Q_o+Q_v|^2 \Lambda_{4,4} \Lambda_{2,2}[^0_0] + |Q_o-Q_v|^2
|X_o-X_v|^2 \Lambda_{2,2}[^0_1] \} \ee Massless states are
contributed by $|Q_o X_o|^2$ and $|Q_vX_o|^2$, that produce in all
12 massless vectors ($G_{\mu,4\pm i 5}$, $B^{(2)}_{\mu,4\pm i 5}$,
$C^{(2)}_{\mu,4\pm i 5}$, $C^{(4)u}_{\mu,4\pm i 5}$), while the 16
twisted sectors \be \cT^{\cN=4}_t = {16\over 2} \{ |Q_s+Q_c|^2
|X_s+X_c|^2 \Lambda_{2,2}[^1_0] + |Q_s-Q_c|^2 |X_s-X_c|^2
\Lambda_{2,2}[^1_1] \} \ee only contribute massive states because
of the shift.

The second step consists in a non geometric (say Left-) projection
combined with a shift along the orthogonal directions. The
partition function in the untwisted sector reads \bea
&&\cT^{\cN=3}_u = {1\over 2} \{ |Q|^2 \Lambda^{(4,5)}_{2,2}[^0_0]
\Lambda^{(6,7)}_{2,2}\Lambda^{(8,9)}_{2,2}+
|Q^{(4,5)}_o-Q^{(4,5)}_v|^2
\Lambda^{(4,5)}_{2,2}[^0_1] + \\
&&[(Q_o-Q_v)^{(6,7)}
\Lambda^{(6,7)}_{2,2}[^0_1]\Lambda^{(8,9)R}_{2,2} +
(Q_o-Q_v)^{(8,9)}
\Lambda^{(6,7)R}_{2,2}\Lambda^{(8,9)}_{2,2}[^0_1]]\bar{Q}
\Lambda^{(4,5)R}_{2,2}[^0_0]\} \nn \eea where the superscript
$(I,I+1)$ denote the `untwisted' directions in a given sector.

Twisted sectors only contribute massive states. The only massless
states thus arise from \bea (V_2 - S_2 - C_2)\times (\bar{V}_2 + 2
\bar{O}_2 - 2\bar{S}_2 -
2 \bar{C}_2) \rightarrow  \nn \\
(g+b+\phi)_{_{NS-NS}} + 2 A_{_{NS-NS}} + 4 \phi_{_{R-R}} + 2
A_{_{R-R}} - Fermi \eea and \be (2O_2 - S_2 - C_2) \times (4
\bar{O}_2 - 2\bar{S}_2 - 2 \bar{C}_2) \rightarrow
\\ 8 \phi_{_{NS-NS}} + 4 \phi_{_{R-R}} + 2 A_{_{R-R}} -
Fermi \ee and correspond to  $\cN=3$ supergravity coupled to three
vector multiplets, as anticipated. In all, there are 2 NS-NS
vectors and 4 R-R vectors. The scalar moduli parameterize
$\cM_{\cN=3}^{D=4} = SU(3,3)/SU(3)\times SU(3)\times U(1)$. Vector
fields together with their magnetic dual transform according to
the (complex) ${\bf 6}$ of $SU(3,3)$.

The surviving NS-NS charges are $p_{_R}^a$ with $a=4,5$ and their
magnetic duals $\hat{P}_{_R}^a$.

The R-R charges are the geometric charges of the parent $\cN
=4=2_{_L}+2_{_R}$ theory left invariant by the double T-duality
projection. Starting with \be q_1^a \quad , \quad q_5^a \quad ,
\quad q_3^{aij} \ee one finds that the T-duality invariant
combinations are simply \be q_1^a+{1\over 3!} \varepsilon^a_{bij}
q_3^{bij} \quad , \quad q_3^{aij} + {1\over 3!}
\varepsilon^a_{bkl} q_5^{bijkl}\ee

In $\cN =3$ supergravity, there are no 1/2 BPS particle states.
One can consider 1/3 BPS states, which descend from 1/2 or 1/4 BPS
states in the `parent' $\cN =4$ theory \cite{Kounnas:1997hi,
Andrianopoli:2002rm}. Bound-states of D-branes carrying the one of
the above charges are 1/3 BPS the same is true for states carrying
KK momenta, windings or their magnetic duals. Notice however that
the gravitino which becomes massive in the breaking of $\cN=4$ to
$\cN=3$ belongs to a long multiplet.

In \cite{Ferrara:1989nm} a complete classification of ``minimal''
$\cN =3$ superstring solutions was given. Depending on the choice
of fermionic sets, there are four classes with $3+4K$ matter
vector multiplets, with $K=0,1,2$, that give rise to some eleven
sub-classes. Moreover with an extra chiral projection (ie
splitting the geometric K3 into two chiral $\Z_2$) one can get
models with $1+2K$ matter vector multiplets, with $K=0,1,2$. In
particular a model with only one vector multiplet, containing
three complex scalars, is possible.

Another construction attributed to Narain in
\cite{Ferrara:1989nm}, is an asymmetric $\Z_3$ projection with
$\theta = (\omega_3,\omega_3, \omega_3; 1, \omega_3,
\omega_3^{-1})$ acting on the lattice of $SU(3)^3$.

\subsection{$\cN =2= 1_{_L}+1_{_R}$ case}

We will be very brief about this case since, choosing geometric
projections, it includes widely studied Calabi-Yau
compactifications of Type II superstrings. In particular the
geometric $T^6/\Z_2\times \Z_2$ orbifold can be resolved to a CY
3-fold with Hodge numbers $h_{11} = 51$ and $h_{21} = 3$ or its
mirror, related to one another by the choice of discrete torsion.

The $\cN =2=1_{_L} + 1_{_R}$ case can also be obtained by
non-geometric $\Z_2^L \times \Z_2^L \times \Z_2^R \times \Z_2^R$
projections, \ie by adding a further $\Z_2^R$ projection onto the
$\cN =3=1_{_L} + 2_{_R}$ case. One can either perform a specular
projection that leads to a non geometric but L-R symmetric model
that allows for further $\Omega$ projection \cite{wip} or perform
a different R projection leading to a genuinely L-R asymmetric
model. Alternatively one can consider $\Z_2 \times \Z_2$ with L-R
asymmetric shifts acting on both K-K momenta and windings. With
generic choices of the shifts, there are no massless twisted
states and no NS-NS massless vectors. There are however 4 R-R
vectors that together with their magnetic duals couple to the
eight different kinds of D3-branes one can wrap around the
internal `3-cycles'\footnote{We put 3-cycles in quotes since the
background is non-geometric.}.

Another possibility is to consider $\Z_3^L \times \Z_3^R$ with L-R
asymmetric shifts breaking directly $\cN =8=4_{_L} + 4_{_R}$ to
$\cN =2=1_{_L} + 1_{_R}$. In this case only the R-R graviphoton
survives in the untwisted sector and together with its magnetic
dual it couples to D3-branes wrapped around the `3-cycles' dual to
the holomorphic and anti-holomorphic 3-forms.

Another interest class are the `magic' supergravities recently
constructed in \cite{Bianchi:2007va, Dolivet:2007sz}.

\section{Open string excitations}

A convenient vantage point for identifying the relevant open
string excitations is to use the boundary state formalism
\cite{Abouelsaood:1986gd, Callan:1987px}. In this formalism
D-branes are represented as coherent or rather `squeezed' states
of closed string harmonic oscillators. The transverse
boundary-to-boundary (cylinder) amplitude reads \be \tilde{\cal
A}_{ab} = \langle B_a|\exp(-\pi \ell {\cal H}_{cl}) |B_b\rangle
\quad , \ee where $a,b$ label the (different) boundary states.

For (obliquely) magnetized D9-branes \cite{Abouelsaood:1986gd,
Bachas:1995ik, Antoniadis:2004pp, Bianchi:2005yz,
Antoniadis:2005nu, Bianchi:2005sa, Di Vecchia:2006gg} the
contribution of the bosonic coordinates reads \be
|B_a\rangle^{(X)} = \sqrt{\det ({\cal G}_a + {\cal F}_a)} \exp (-
\sum_{n>o} a^i_{-n} R_{ij}(F_a) \tilde{a}^j_{-n}) |0_a\rangle \ee
where \be R_a = {1 - F_a \over 1 +F_a} \ee is the relative
`rotation' between Left and Right movers induced by the internal
magnetic field. The zero-mode contribution is implicit in
$|0_a\rangle $ and consists in a sum over all $p_{_L} = - R_a
p_{_R}$. For non-compact directions $p_{_L} = - p_{_R}$.

The contribution of the fermionic coordinates to the boundary
state is somewhat subtler. In the NS-NS sector, there are no
fermionic zero-modes and one has \be |B_a, \eta
\rangle_{_{NS-NS}}^{(\psi)} = \exp (i\eta\sum_{n\ge 1/2}
\psi^i_{-n} R_{ij}(F_a) \tilde{\psi}^j_{-n}) |\eta\rangle \ee
where $\eta=\pm$ stands for possible GSO projections and choice of
superghost picture. In the R-R sector, fermions admit zero-modes,
that cancel the Born-Infeld action and replace it with the
Wess-Zumino coupling \bea &&|B_a, \eta
\rangle_{_{R-R}}^{(\psi,\b,\g)} = {1 \over \sqrt{\det ({\cal G}_a
+ {\cal F}_a)}} \exp (i\eta \sum_{n>0} \tilde{\psi}^i_{-n}
R_{ij}(F_a) \psi^j_{-n}) |0_a, \eta\rangle \eea where \be |0_a,
\eta\rangle = {\cal U}_{A\tilde{B}}(F_a) |A,\tilde{B}\rangle \ee
with \be {\cal U}_{A\tilde{B}}(F_a) = \left[{\rm AExp} (-
F^a_{ij}\Gamma^{ij}/2) C\Gamma_{11} {1 + i\eta \Gamma_{11}\over 1
+i\eta}\right]_{A\tilde{B}}\quad . \ee where ${\rm AExp}$ means
antisymmetrization of the vector indices of the $\Gamma$ matrices.

Magnetized D-branes in L-R symmetric orbifolds have been
considered in \cite{Blumenhagen:2000wh, Angelantonj:2000hi}. The
case of $g_{_L}= g_{_R}^{-1}$, where $g_{_L}$ and $g_{_R}$ are the
generators of the orbifold group $\Gamma_{_{L\neq R}}$ on L- and
R- movers, was discussed in \cite{Blumenhagen:2000fp,
Angelantonj:2000xf, Gutperle:2000bf} and is equivalent to choosing
a different (diagonal) modular invariant torus partition function.
We would like to generalize the analysis to the case of genuinely
L-R asymmetric orbifolds in which $g_{_L} \neq g_{_R}$. The action
of a L-R asymmetric rotation on the superstring coordinates imply
\be g_{_L} g_{_R}|B, F_a\rangle = |B, F'_a\rangle\ee where the
action of the shift on the bosonic zero-modes is understood and
the `transformed' magnetic field $F_a'$ is such that \be R(F'_a) =
R(g_{_L}) R(F_a) R^t(g_{_R}) \ee For a
$Z^L_{N_{_L}}\times{Z^R_{N_{_R}}}$ action, invariant boundary
states would then be of the form \bea |B, F\rangle_g &=& {1\over
\sqrt{N_{_L} N_{_R}}} \left(1 + g_{_L} + g_{_R} + .... +
g^{N_{_L}-1}_{_L} g^{N_{_R}-1}_{_R}\right) |B, F\rangle = \nn \\
&=& {1\over \sqrt{N_{_L} N_{_R}}} \sum_{l,r} |B,F_{(l,r)}\rangle
\eea where the `induced' magnetic field $F_{(l,r)}$ is determined
by the condition \be R(F_{(l,r)}) = R(g^l_{_L}) R(F) R^t(g^r_{_R})
\ee

Computing the amplitude \be \tilde{\cal A}(F)_g = {}_g\langle B,
F|\exp(-\pi \ell {\cal H}_{cl}) |B, F\rangle_g  \quad , \ee one
can easily identify the couplings of the invariant magnetized
D-brane to the closed string states of the L-R asymmetric orbifold
and check the BPS no-force condition. Performing a modular $S$
transformation one can then find the open string excitations. The
typical term would be of the form \be \cA_{g, h} = \Lambda(g,h)
\cI(g,h) \sum_\a c_\a^{^{GSO}}{\vartheta_\a(0)\over \eta^3}
\prod_I {\vartheta_\a(\epsilon_I(g,h)\tau)\over
\vartheta_1(\epsilon(g,h)\tau)} \ee where $g=g_{_L} g_{_R}$ ,
$h=h_{_L} h_{_R}$, $\Lambda(g,h)$ is the lattice invariant under
$gh=g_{_L} h_{_L} g_{_R} h_{_R}$, while $\cI(g,h)$ is the
`intersection' number counting the invariant discrete zero-modes
and finally $\epsilon_I(g,h)$ are related to the eigenvalues of
$gh=g_{_L} h_{_L} g_{_R} h_{_R} \rightarrow
diag(e^{2i\epsilon_I(g,h)})$. The BPS condition in the transverse
channel translates into the supersymmetry condition $\sum_I
\epsilon_I = 0$ (mod 1).

Requiring integer multiplicities may put some additional
constraints on the choice of $F$ and of the phases in the
projection \cite{Blumenhagen:2000fp, Angelantonj:2000xf,
Gutperle:2000bf}. Due to the relative `rotations' among the
various components in the invariant boundary state, both `twisted'
(non integer moded) and `untwisted' (integer moded) open strings
will appear in the spectrum of a single D-brane bound state. For
L-R symmetric non geometric Type I compactifications this has
already been observed in \cite{Brunner:1999fj, Bianchi:1999uq,
Gaberdiel:2002jr} and was anyway implicit in the systematic
construction of \cite{Bianchi:1988fr, Bianchi:1989du}

Let us illustrate the above procedure for the bound-state of
D-branes in the $\cN=5$ model obtained by asymmetric $\Z_3$
projection on the torus of $SU(3)^3$ with shift along $v =
\alpha_1/3$ which is not a lattice vector, while $3v$ is. Notice
that prior to twists and shifts there are 27 boundary states
associated to the 27 conjugacy classes of $SU(3)^3$. Let us denote
these by $\vec{r}=(r_1,r_2,r_3)$ with $r_i=0,1,2(\equiv -1 {\rm
mod 3})$. The direct channel annulus amplitude between any two of
these reads \be \cA_{\vec{r},\vec{s}} = N_{\vec{r},
\vec{s}}^{\vec{t}} \cX_{\vec{t}} \ee where $\cX_{\vec{t}} =
(V_8-S_8) \chi_{t_1}\chi_{t_2}\chi_{t_3}$ denote the
super-characters, the only massless one being $\cX_{\vec{0}} =
(V_8-S_8) \chi_{0}\chi_{0}\chi_{0}$. The fusion rule coefficients
$N_{\vec{r}, \vec{s}}^{\vec{t}}$ simply reflect the $\Z_3^3$
selection rules of the center of $SU(3)^3$. The transverse channel
amplitude is \be \widetilde\cA_{\vec{r},\vec{s}} =
B_{\vec{r}}^{\vec{t}} B_{\vec{s}}^{\vec{t}} \cX_{\vec{t}} \ee
where the boundary reflection coefficients are simply given by \be
B_{\vec{s}}^{\vec{t}} = {S_{\vec{s}}^{\vec{t}} \over
\sqrt{S_{\vec{0}}^{\vec{t}}}} \rightarrow {\omega_3^{\vec{s}\cdot
\vec{t}}\over \sqrt[4]{3^3}} \ee

In the language of magnetized branes the three boundary states per
each $T_{SU(3)}^2$ correspond to branes with magnetic quantum
number $(n,m) = (1,0), (-1,1), (0,-1)$, which are T-dual to
rotated branes. Clearly one has \be \cA^{SU(3)}_{a,b} = \langle
B_a| q^H |B_b\rangle = N_{ab}{}^c \chi_c^{SU(3)} \ee

We are now ready to discuss the effect of twists and shifts.
Since, in this case, we can represent the original boundary states
as magnetized (or rotated) brane states, the surviving `regular'
bound-states are simply given by \be |B_a\rangle_{\Z_3^{L\neq R}}
= {1\over \sqrt{3}} \left(|B_a\rangle + \theta_{_L} \sigma_{_R}
|B_a\rangle + \theta^2_{_L} \sigma^2_{_R} |B_a\rangle\right) \ee
or more explicitly \bea &&|B_a\rangle_{\Z_3^{L\neq R}} = {1\over
\sqrt{3}}
[\sum_{{^{p_{_L} =}_{ - R_a p_{_R}}}} |B; R_a, p_{L|R}\rangle \\
&& + \sum_{ {^{R_{_L}p_{_L} =}_{ - R_a p_{_R}}}} e^{2\pi i v_{_R}
p_{_R}}|B; R_{_L} R_a, p_{L|R}\rangle + \sum_{ {^{R^{-1}_{_L}
p_{_L} =}_{ - R_a p_{_R}}}} e^{-2\pi i v_{_R} p_{_R}}|B;
R^{-1}_{_L} R_a, p_{L|R}\rangle ] \nn \eea where $R_{_L}$ is the
(chiral) rotation of $2\pi/3$ and $v_{_R}$ parameterizes the shift
of order three.

Computing the self-overlap of an invariant boundary state, yields
the annulus amplitudes in the closed string `tree-level' channel
where the coupling to the 4 surviving R-R graviphotons can be
extracted. Performing a modular $S$ transformation yields the open
string `loop' channel that reads \be \cA_{\Z_3^{L\neq R}} =
{1\over 6} \sum_{a,b \in \Z_3} \Lambda_{(a,b)} \cI_{(a,b)} \sum_\a
c_\a^{^{GSO}} {\vartheta_\a(0)\over \eta^3} \prod_I
{\vartheta_\a(a\tau + b)\over \vartheta_1(a\tau + b)} \ee where
$\Lambda_{(a,b)}$ is the shifted lattice sum, appearing only for
$a=b=0$ and depending on the choice of boundary state, and
$\cI_{(a,b)}$ are multiplicities representing the number of
`chiral' fixed points, \ie $\cI_{(0,0)}=1$, $\cI_{(\pm 1,b)}=3$
and $\cI_{(0,\pm 1)}=3$.

In addition to the `regular' bound-states of D-branes, the $\cN=5$
model obtained by asymmetric $\Z_3$ projection on the torus of
$SU(3)^3$ should admit `fractional' ones that would couple to
massive twisted fields.

The above procedure can be straightforwardly generalized and
implemented in L-R asymmetric orbifolds of `geometric' superstring
vacua whenever D-brane boundary states are known or computable.

\section{Outlook}

In this note, we have discussed bound-states of D-branes in
genuinely L-R asymmetric and thus non-geometric Type II vacuum
configurations with extended supersymmetry. These L-R asymmetric
D-branes couple to the R-R graviphotons surviving in the massless
spectrum. We have also shown that they preserve a fraction of the
supersymmetries of the background and described a procedure to
identify the relevant open string excitations thereof. Boundary
state techniques apply to general (rational) CFT and thus one can
envisage the possibility of extending the present analysis to
Gepner models \cite{Angelantonj:1996mw, Recknagel:1997sb,
Blumenhagen:1998tj, Dijkstra:2004cc} or other abstract CFT's
combined in a L-R asymmetric fashion, after some twist and shift.

In addition to being interesting in their own right, since L-R
asymmetric backgrounds, though promising, are largely unexplored
and above all since their very existence looks rather
counter-intuitive, L-R asymmetric D-branes may find concrete
applications in Black Hole attractor solutions and their
microstate counting. In particular, the boundary state description
we have adopted for the identification of the open string
excitations can be easily adapted to cases with several stacks of
`intersecting' L-R asymmetric D-branes combined with (surviving)
fundamental string windings and KK momenta. Moreover the analogue
of fractional D-branes should be present also in genuinely L-R
asymmetric backgrounds. Some simple instances have already been
proposed \cite{Gaberdiel:2002jr, Kawai:2007qd}.

In this note we have only considered bound-states of D-branes that
couple to R-R graviphotons and are thus point-like objects
(particle states) along the non-compact directions. Since the L-R
asymmetric projections only act in the compact `directions', it is
almost trivial to construct bound-states of D-branes that are
extended along some non-compact direction. In particular,
invariant bound states of (magnetized) D9-branes and other lower
dimensional branes are an obvious possibility. The existence of
invariant bound-states of `intersecting' or differently magnetized
D-branes invading the non-compact spacetime directions calls for
the existence of L-R asymmetric $\Omega$-planes with opposite R-R
charge and `tension'. Indeed given the similarity between crosscap
and boundary states one might be tempted to simply consider $g$
invariant combinations of $\Omega$-planes.  If the naive guess is
correct as for the L-R asymmetric D-branes one could start
building an entire new class of `unoriented' vacuum configurations
where the L-R asymmetric twists and shifts fix many if not all of
the closed string moduli \cite{Silverstein:2001xn} and the
`intersecting' L-R asymmetric unoriented D-brane account for the
massless gauge and matter fields.

\section*{Acknowledgements}

The author would like to acknowledge P.~Anastasopoulos,
F.~Marchesano, J.F.~Morales, G.~Pradisi, N.~Prezas, A.~Sen and
especially S.~Ferrara for illuminating discussions and to thank
A.~Sagnotti and C.~Angelantonj for valuable comments on the
manuscript. The work has been supported in part by the European
Community Human Potential Program under contract
MRTN-CT-2004-512194, by the INTAS grant 03-516346, by MIUR-COFIN
2003-023852, and by NATO PST.CLG.978785.

\end{document}